\documentclass[10pt]{article}
\usepackage{epsf,latexsym,rotate}
\usepackage{amsfonts,amssymb} %%amsmath
\epsfverbosetrue
\usepackage{graphicx}

\newcommand{\id}{\mathrm{id}}

\newcommand{\double}[1]{\mathbb{#1}}

\newcommand{\rr}{\double{R}}

\newcommand{\ttt}{{\rm Tr}}
\newcommand{\tr}{{\rm tr}}

\newcommand{\rank}{\rm{rank}}

\newcommand{\bb}{\begin{eqnarray}}
\newcommand{\ee}{\end{eqnarray}}
\newcommand{\eee}{\nonumber\end{eqnarray}}

\begin{document}

\font\twelve=cmbx10 at 13pt
\font\eightrm=cmr8

\thispagestyle{empty}

\begin{center}

Institut f\"ur Mathematik  $^1$ \\ Universit\"at Potsdam
\\ Am Neuen Palais 10 \\14469 Potsdam \\ Germany\\

\vspace{2cm}

{\Large\textbf{The Spectral Action for Dirac Operators  \\ 
with skew-symmetric Torsion}} \\

\vspace{1.5cm}

{\large Florian Hanisch$^{1,2}$, Frank Pf\"aff\-le$^{1,3}$ \& Christoph A. 
Stephan$^{1,4}$}

\vspace{2cm}

{\large\textbf{Abstract}}
\end{center}
We derive a formula for the gravitational part of the 
spectral action for Dirac operators
on 4-dimensional manifolds with totally anti-symmetric torsion.
We find that the torsion becomes dynamical and couples to
the traceless part of the Riemann curvature tensor. 
Finally we deduce the Lagrangian for the Standard Model
of particle physics in presence of torsion from the 
Chamseddine-Connes Dirac operator.

\vspace{2cm}

\noindent
PACS-92: 11.15 Gauge field theories\\
MSC-91: 81T13 Yang-Mills and other gauge theories

\vskip 1truecm

%\noindent CPT-Pxx-2007\\
\noindent \\

\vspace{1.5cm}
\noindent $^3$ fhanisch@uni-potsdam.de\\
\noindent $^3$ pfaeffle@math.uni-potsdam.de\\
\noindent $^4$ christophstephan@gmx.de\\

\newpage

\subsection*{Introduction}

In classical Relativity one presumes that the gravitational 
degrees of freedom are encoded only in the choice of the metric and
one can therefore restrict to  Levi-Civita connections.
This is usually justified  by Einstein-Cartan theory because the critical
points of the corresponding action are space-times with
torsion free connections. 

If we want to incorporate  the other forces
of Nature (i.e. the electro-weak and the strong force) and the 
known fermions (Leptons and Quarks) we have to consider 
unified theories.
A geometrical approach to such a unification is offered by
Connes' noncommutative geometry \cite{con}. The natural action to
consider in this framework is given by the spectral action 
principle \cite{cc}.

In the language of noncommutative geometry any geometry is encoded by a
spectral triple which consists of an algebra, a module over this algebra and 
a  first order 
differential operator. Here one should picture one of the simplest
examples given by compact Riemannian spin manifolds: the
algebra is formed by the smooth functions, the module is given by
the square-integrable spinor fields and the first order differential
operator is the classical Dirac operator. To reconstruct 
the manifold one needs further ingredients (a real structure,
a chirality operator)  \cite{con}. We will omit any
details  here, since the only relevant objects for this article turn out
to be classical twisted Dirac operators. 

In combination with product geometries based on spectral triples
for Riemannian manifolds and finite geometries one even finds
a conceptual explanation  for the Standard Model of particle physics \cite{mcc}. 
Still, the relevant Dirac operators are classical twisted Dirac operators
and the module consists of  twisted spinor fields.
For these so called almost-commutative geometries,  
the spectral action  precisely predicts the Lagrangian of
corresponding Einstein-Yang-Mills-Higgs model.
It has also successfully predicted scale invariant Lagrangians with
dilaton fields \cite{ccdil}, and  quantum gravity boundary
terms \cite{ccbound} in the case of manifolds with boundary.

In this paper we take the spectral geometric point of view, in the
sense that we are only dealing with classical objects where the
noncommutative input resides in the exact structure of the twist
bundle.
We expand the known calculations of the spectral action (with and
without the standard model) by implementing skew-symmetric torsion 
into the relevant twisted Dirac operators. As a result we find that in this context
the Einstein-Cartan theory does not apply. One finds an action
that suggests critical points with nonzero torsion which might even
be dynamical.

\subsection*{The spectral action for pure gravity with torsion}

In this first section we concern ourselves  only with the gravitational part of the spectral action
for a 4-dimensional closed Riemannian manifolds $M$ with spin structures. 
Let us now briefly recall the basic notions of connections and Dirac
operators with torsion. Each connection on the tangent
bundle of a manifold can be written as a sum of the Levi-Cevita connection 
$\nabla^{LC}$ and a $(2,1)$-tensor field $A$, i.e.\ $
\nabla_X Y = \nabla_X^{LC} Y + A(X,Y)$.
For such a general connection $\nabla$ the  torsion 
3-form is by definition $T(X,Y,Z) = \langle \nabla_X Y - \nabla_Y X
- [X,Y],Z \rangle$.

The connection $\nabla$ is compatible with the Riemannian metric 
$\langle \cdot, \cdot \rangle$ and
has the same geodesics as $\nabla^{LC}$ if and only if 
$A(X,Y,Z)= \langle A(X,Y), Z \rangle$ is totally anti-symmetric.
We will only consider this case.
Then the torsion of $\nabla$ is given by $T=2A$, and hence
\bb
\nabla_X Y = \nabla_X^{LC} Y + \frac{1}{2}T(X,Y,\cdot)^\#.
\label{zero}
\ee
Note the $T(X,Y,\cdot)^\#$ equals the vector valued torsion
2-form of E.\ Cartan, which is defined as the exterior covariant derivative
of the soldering form.

Since we assume that the manifold carries a spin structure, we
can consider spinor fields $\psi$ and the  spin connection induced by
$\nabla$ can be expressed as
\bb
\nabla_X \psi = \nabla^{LC}_X \psi + \frac{1}{4} ( X \lrcorner T) \cdot \psi.
\label{tconnection}
\ee
Here $( X \lrcorner T) \cdot$ denotes Clifford multiplication\footnote{Here we
use the Clifford relations $X\cdot Y + Y \cdot X = -2 \langle X,Y \rangle$ for tangent 
vectors $X,Y$. Any 2-form $X^\flat \wedge Y^\flat$ acts as $\frac{1}{2}
X\cdot Y \cdot$ on the spinor module. } by the 2-form
$T(X,\cdot,\cdot)$. This spin connection yields a Dirac operator $D$ which
one can write as $D \psi = \sum_i e_i \cdot \nabla_{e_i} \psi$, for
any orthonormal frame $e_i$.

From \cite[Thm.\ 6.2]{af} we deduce the Bochner form of the square of this Dirac operator:
\bb
D^2 = \Delta + \frac{3}{4} dT + \frac{1}{4} R - \frac{9}{8} T_0^2,
\label{DD}
\ee
where $\Delta$ is the Laplacian associated to the spin connection
\bb
\widetilde{\nabla}_X \psi = \nabla^{LC}_X \psi + \frac{3}{4} (X \lrcorner T)
\cdot \psi,  
\label{drittelconn}
\ee
$dT$ is the exterior differential of the 3-form $T$,
$R$ is the scalar curvature of the Riemannian manifold (in our convention
spheres have positive curvature, i.e.\ $R=12$ for the 4-dimensional sphere)
and $T_0^2 = \frac{1}{6} \sum_{i,j=1}^n \| T(e_i,e_j,\cdot)^\# \|^2$.

For the Dirac operator $D$ we will calculate the bosonic part of the
spectral action. 
It is defined to be the
number of Eigenvalues of   
$D$ in the interval $[ -\Lambda,  \Lambda ]$
with $\Lambda \in \rr^+$. In \cite{cc} it is expressed  as
\bb
I = \tr \, F \left( \frac{D^2}{\Lambda^2} \right) 
\nonumber
\ee
Here $\tr$ denotes the operator trace in the Hilbert space of
$L^2$-spinor fields, 
and $F:\mathbb{R}^+ \to \mathbb{R}^+$ is a cut-off function with support in
the interval $[0,+1]$ which is constant near the origin. Here we follow
the notation of \cite{cc2}. 

For $t \rightarrow 0$ one has the heat trace asymptotics \cite{gilk}
\bb
\tr \left( e^{-t \, D^2} \right) \sim  \sum_{n\geq 0} t^{n-2} a_{2n} (D^2)
\nonumber
\ee 
One uses the Seeley-deWitt coefficients $a_{2n}(D^2)$ and $t= \Lambda^{-2}$ 
to obtain an asymptotics for the spectral action \cite{cc,nvw}
\bb
I = \tr \, F \left( \frac{D^2}{\Lambda^2} \right) \sim 
\Lambda^4 \, F_4 \, a_0 (D^2) + \Lambda^2 \, F_2 \, a_2(D^2)  
+ \Lambda^0 \, F_0 \, a_4(D^2) \quad \mbox{as } \;\; \Lambda \to \infty 
\label{specact}
\ee
with the first three moments of the cut-off function which are given
by
$F_4 = \int_0^\infty s \cdot F(s) \, ds$, $F_2 = \int_0^\infty F(s) \, ds$ and $F_0 = F(0)$.
Note that these moments are independent of the geometry of the manifold.

Setting $E = - \frac{3}{4} dT - \frac{1}{4} R + \frac{9}{8} T_0^2$, we get
$D^2 = \Delta - E$ from (\ref{DD}). We use \cite[Thm.\ 4.1.6]{gilk} to obtain the 
first three coefficients of the heat trace asymptotics:
\bb
a_0 (D^2) &=&   \frac{1}{4 \pi^2}  \int_M dvol \\
a_2 (D^2) &=&   \frac{1}{96 \pi^2}  \int_M ( 6 \, \tr (E) + 4 R) \, dvol \\
a_4 (D^2) &=&   \frac{1}{5760 \pi^2}  \int_M  \Big(  \tr \left( 60 \, \Delta E +60 R E + 180 E^2 +
30 \, \Omega_{ij} \Omega_{ij}\right) \nonumber \\
&& \qquad \qquad \quad + 48 \Delta^{LC} R + 20 R^2 - 8 \|Ric\|^2 + 8 \|Riem\|^2 \Big) \, dvol 
\nonumber
\ee
Here $Ric$ and $Riem$ denote the Ricci curvature and the Riemannian curvature
tensors of the metric, and $\Omega_{ij}=\widetilde{\nabla}_{e_i}  \widetilde{\nabla}_{e_j}
-\widetilde{\nabla}_{e_j} \widetilde{\nabla}_{e_i} - \widetilde{\nabla}_{[e_i,e_j]} $ is
the curvature of $\widetilde{\nabla}$.

We evaluate the $a_i(D^2)$ above and take into account that  
 $\tr (dT) = 0$ due to Clifford relations and cyclicity of the trace.
For the first two coefficients we get
\bb
a_0 (D^2) =   \frac{1}{4 \pi^2}  \int_M dvol, \qquad
a_2 (D^2) =   \frac{1}{16 \pi^2} \int_M \left(  \frac{9}{2}  T_0^2 - \frac{1}{3} R \right) dvol.
\nonumber
\ee
For  $a_4(D^2)$ we use that $\tr (\Delta E) = \Delta^{LC} \tr (E)$ and $\Delta^{LC} R$
vanish after integration over the closed manifold $M$.
\bb
a_4(D^2) = \frac{1}{16 \pi^2} \int_M &\Big(& \frac{1}{72} R^2 - \frac{1}{45} \|Ric\|^2
+  \frac{1}{45} \|Riem\|^2 \nonumber \\ 
&&- \frac{3}{8} R T_0^2 + \frac{9}{8} \|dT\|^2 + \frac{81}{32}
(T_0^2)^2 + \frac{1}{12} \sum_{i,j} \tr ( \Omega_{ij} \Omega_{ij}) \Big) \, dvol
\nonumber
\ee
Similar calculations have been done in \cite{OBU}. 
For the curvature $\Omega_{ij}$ of the connection $\widetilde{\nabla}$ we proceed
by computing
\bb
\Omega_{ij} = \sum_{a,b} &\Big(& \frac{1}{4} \langle R(e_i,e_j)e_a,e_b \rangle
+ \frac{3}{8} a(\nabla T) (e_i,e_j,e_a,e_b) \nonumber \\
&&+ \frac{9}{16} \sum_c \big(
T (e_i,e_c,e_a)T (e_j,e_c,e_b) - T (e_j,e_c,e_a)T (e_i,e_c,e_b)
\big) \Big)e_a e_b, 
\nonumber
\ee
where $a(\nabla T)$ denotes the anti-symmetrisation in the first two entries of $\nabla T$,
i.e.\ $a(\nabla T) (e_i,e_j,e_a,e_b) $ \linebreak $= \nabla_{e_i} T (e_j,e_a,e_b) - \nabla_{e_j} T (e_i,e_a,e_b)$. Using the identity $\tr (e_k e_l e_s e_t) = 4 (\delta_{ls} \delta_{kt}  - \delta_{lt} 
\delta_{ks}) $ we obtain
\bb
\sum_{i,j} \tr ( \Omega_{ij} \Omega_{ij}) = - 8  \, \sum_{ {i\neq j} \atop{a,b}}
\Big( \frac{1}{4}\langle R (e_i,e_j)e_a,e_b \rangle + \frac{3}{8} a(\nabla T)
(e_i,e_j,e_a,e_b) 
+ \frac{9}{8} c(T) (e_i,e_j,e_a,e_b)\Big)^2,   
\nonumber
\ee
where $c(T) (e_i,e_j,e_a,e_b)  = \sum_c T(e_i,e_c,e_a) T(e_j,e_c,e_b)$. 
This term equals the square of a norm in the space of $(4,0)$-tensors.
We use representation theory \cite[Chap.~4]{SAL} of $O(4)$ to 
decompose these tensors into irreducible components. 
We note that $c(T)$ is a formal curvature tensor after interchanging the second and the third entry and hence we may write
$Riem,c(T) \in \mathbb{R} \oplus \mathrm{Sym}_0^2 \oplus \mathrm{Weyl}$.
Here we consider $ \mathbb{R} \oplus \mathrm{Sym}_0^2 \oplus \mathrm{Weyl}
\subset \mathrm{Sym}^2 ( \Lambda^2)$.
Furthermore $a: \Lambda^1 \otimes \Lambda^3 \to a(\Lambda^1 \otimes \Lambda^3)
\subset \Lambda^2 \otimes \Lambda^2$ is an isomorphism of $O(4)$-representations.
The image splits into $a(\Lambda^1 \otimes \Lambda^3) = \Lambda^4
\oplus \Lambda^2 \oplus \mathrm{Sym}^2_0$, where $\Lambda^4 \subset 
\mathrm{Sym}^2( \Lambda^2)$ and $ \Lambda^2 \oplus \mathrm{Sym}^2_0
\subset \Lambda^2 (\Lambda^2)$.
As the above decompositions are orthogonal, we conclude that 
$c(T) \perp a(\nabla T) $ and $Riem \perp a(\nabla T)$ in the space of $(4,0)$-tensors.

After identification by the above isomorphisms, this yields
that the norm of the $\Lambda^4$-component of $a(\nabla T)$  equals 
$\sqrt{6} \|dT\|$ and norm of  the $\Lambda^2$-component is $2 \|d^* T\|$, 
The norm of the remaining
component in $\mathrm{Sym}_0^2$ is denoted by $2 \|sym_0^2 (\nabla T)\|$.
We compute:
\bb
\frac{1}{12} \sum_{i,j} \tr ( \Omega_{ij} \Omega_{ij}) =
 - \frac{1}{24} \| Riem \|^2 - \frac{3}{8} \| d^*T\|^2
 -\frac{9}{16} \| dT\|^2 - \frac{3}{8} \|sym_0^2 (\nabla T)\|^2
 - \frac{27}{32} \| c(T) \|^2 - \frac{3}{8} P(T),
\nonumber
\ee
where we abbreviate 
\bb
P(T)&:=&  \sum_{ {i\neq j} \atop{a,b,c}}   
\langle R (e_i,e_j)e_a,e_b \rangle  T(e_i,e_c,e_a) T(e_j,e_c,e_b).
\nonumber
\ee
Once more we insert the Ricci decomposition of the  curvature tensor into 
the  scalar curvature, the  
traceless Ricci tensor  and the Weyl tensor $W$ and obtain
\bb 
P(T) &=& - R T_0^2 -  \sum_{ {i\neq j} \atop{a,c}} Ric(e_j,e_a) T(e_i,e_c,e_a)
T(e_j,e_c,e_i) \nonumber \\  
&&+ \sum_{ {i\neq j} \atop{a,b,c}}   
\langle W (e_i,e_j)e_a,e_b \rangle  T(e_i,e_c,e_a) T(e_j,e_c,e_b).
\nonumber
\ee
Finally we obtain for the fourth heat coefficient with $R(T) = P(T) + 
R T_0^2$
\bb
a_4(D^2) = \frac{1}{16 \pi^2} \int_M &\Big(& \frac{1}{72} R^2 - \frac{1}{45} \|Ric\|^2
-  \frac{7}{360} \|Riem\|^2  + \frac{81}{32}
(T_0^2)^2  
 - \frac{27}{32} \| c(T) \|^2 \nonumber \\
&&    + \frac{9}{16} \|dT\|^2
 - \frac{3}{8} \| d^*T\|^2
 - \frac{3}{8} \|sym_0^2 (\nabla T)\|^2
  - \frac{3}{8} R(T)
 \Big) \, dvol
\nonumber
\ee
Neglecting the term $a_4$ in the spectral action, we would obtain the classical
Einstein-Cartan-action which has only torsion free critical points upon variation
of metric and torsion 3-forms.
A similar action functional for Dirac operators with totally anti-symmetric
torsion has already been considered in \cite{KW,AT}, where the
authors used the Wodzicki residue as the bosonic action. This 
gives  an action involving only the second Seeley-deWitt coefficient.
 
Considering the full  spectral action (\ref{specact}), which also
includes $a_4(D^2)$, we observe
that the  term $R(T)$ generically
couples torsion and the trace free component of the curvature
tensor. Therefore we expect critical points of the spectral action with
non-zero torsion. Furthermore, due to the derivative terms  of  $T$, 
the torsion becomes dynamical.

\subsection*{The spectral action for the Standard Model  with torsion}

In the noncommutative approach to the Standard Model of particle 
physics,  the fermionic Hilbert space is the product space
of the Hilbert space of $L^2$-sections in the spinor bundle $S$
and a finite dimensional Hilbert space $\mathcal{H}_f$ (called the finite or
internal Hilbert space). The specific particle model is encoded in
$\mathcal{H}_f$. The other important ingredient is a generalised 
Dirac operator $D_\Phi$ acting in the Hilbert space $\mathcal{H} = L^2(M,S) 
\otimes \mathcal{H}_f$, where we follow the notation in \cite{iks}.

On the twisted spinor bundle $S \otimes \mathcal{H}_f$ one considers
a connection $\widehat{\nabla}^{SM} = \nabla \otimes \id_{\mathcal{H}_f}
+ \id_{S} \otimes \nabla^{\mathcal{H}_f}$, where $\nabla$ is a 
connection with skew-symmetric 
torsion as in (\ref{tconnection}) and $ \nabla^{\mathcal{H}_f}$
is a covariant derivative in the trivial bundle $\mathcal{H}_f$ induced 
by gauge fields.\footnote{The Clifford multiplication by a tangent vector $X$ acts as
$X\cdot (\psi \otimes \chi) = (X\cdot \psi) \otimes \chi$. Note that 
the twisted connection $\widehat{\nabla}^{SM}$ is compatible with the 
Clifford multiplication.} 
The associated Dirac operator to $\widehat{\nabla}^{SM}$
is called $D^{\widehat{\nabla}^{SM}}$.
The generalised Dirac operator of the Standard Model
$D_\Phi$ contains  the Higgs boson, Yukawa
couplings, neutrino masses and the CKM-matrix encoded in a field $\Phi$ of 
endomorphisms of $\mathcal{H}_f$.
We follow the conventions of Chamseddine and Connes \cite{cc,mcc} and define 
$D_\Phi$ for  sections $\psi \otimes \chi \, \in \mathcal{H}$ as
\bb
D_\Phi (\psi \otimes \chi)
&=&D^{\widehat{\nabla}^{SM}} (\psi \otimes \chi) +  \gamma_5 \psi \otimes \Phi \, \chi 
\ee 
where $\gamma_5 = e_0 e_1  e_2  e_3$ is the volume element and  $D^{\widehat{\nabla}^{SM}}$ is the twisted Dirac operator (compare
(\ref{DPhi}) in the appendix). We note that $D_\Phi$ is
required to be a self-adjoint operator, consistent with the axioms
of noncommutative geometry \cite{con}. From this one gets restrictions
on $\Phi$, in particular it has to be self-adjoint and compatible with
the real structure $J$ and the chirality operator. We choose the same
$\Phi$ as  Chamseddine and Connes \cite{cc,mcc} since the torsion
does not effect these relations.
  
The bosonic part of the Lagrangian of the Standard Model is obtained
by replacing $D$ by $D_\Phi$ in (\ref{specact}). First we need to calculate
the Bochner formula for the square of  $D_\Phi$.
We use the results from the appendix (e.g. the definition of $\overline \nabla$
(\ref{overline}) 
and the Bochner formula for $D^{\widehat{\nabla}^{SM}}$ (\ref{bochnertwist}))
and get the following Bochner formula
\bb
D_\Phi^2 (\psi \otimes \chi)
&=&\left(D^{\widehat{\nabla}^{SM}}\right)^2 (\psi \otimes \chi) +  
\sum_{i=1}^n \left\{  (\gamma_5 e_i \cdot \nabla_{e_i} \psi ) \otimes \Phi \, \chi  
+ (e_i \cdot \gamma_5 \nabla_{e_i} \psi) \otimes \Phi \, \chi  \right\}
\nonumber \\
&& 
+ \sum_{i=1}^n \left\{  \gamma_5 e_i \cdot  \psi  \otimes \Phi \, 
\nabla_{e_i}^{\mathcal{H}_f} \chi  
+ e_i \cdot \gamma_5  \psi \otimes \nabla_{e_i}^{\mathcal{H}_f} (\Phi \, \chi)  \right\} 
+ (\gamma_5)^2 \psi \otimes (\Phi^2) \chi
\nonumber \\
&=& \left(D^{\widehat{\nabla}^{SM}}\right)^2 (\psi \otimes \chi) -
\sum_{i=1}^n  \gamma_5  e_i \cdot   \psi \otimes [ \nabla_{e_i}^{\mathcal{H}_f} , \Phi] \, \chi 
+  \psi \otimes (\Phi^2) \chi
\nonumber \\
&=& \Delta^{\overline{\nabla}} (\psi \otimes \chi) -E_\Phi (\psi \otimes \chi),
\ee 
where the potential is defined as
\bb
E_\Phi (\psi \otimes \chi) = \widetilde E (\psi \otimes \chi)  
+ \sum_{i=1}^n  \gamma_5  e_i \cdot   \psi \otimes [ \nabla_{e_i}^{\mathcal{H}_f} , \Phi] \, 
\chi  -  \psi \otimes (\Phi^2) \chi
\label{EPhi}
\ee
with $\widetilde E$ as in (\ref{twistpotential}).

We denote the trace on $\mathcal{H}$ and on $\mathcal{H}_f$ as
$\ttt$ and $\tr_f$, respectively (both pointwise and the $L^2$-sense). 
As above $\tr$ is the trace for spinorial
part $S$.
From (\ref{EPhi}) one obtains the trace  
$\ttt(E_\Phi)  = \rank \, \mathcal{H}_f \cdot \tr ( E ) - 4 \, \tr_f (\Phi^2) $,
since the endomorphism $\sum_ {i\ne j} \left(e_i\cdot e_j\cdot\psi\right)\otimes\
\left(\Omega^{\mathcal{H}}_{ij}\;\chi \right)$ in (\ref{Endo}) is skew-symmetric 
and hence traceless.
 
For the  Seeley-deWitt coefficient $a_4(D_\Phi^2)$ we also need to calculate
\bb
(E_\Phi)^2 (\psi \otimes \chi) &=&
\left(E^2\,\psi \right)\otimes\chi
+ \frac{1}{4} \cdot\sum_ {{i \ne j} \atop{k \ne \ell}} \left(e_i\cdot e_j\cdot
e_k \cdot e_\ell \cdot \psi\right)\otimes\left(\Omega^{\mathcal{H}_f}_{ij}
\, \Omega^{\mathcal{H}_f}_{k \ell}\;\chi \right) 
\nonumber \\
&&+ \sum_{i,j=1}^n  \gamma_5  e_i \cdot  \gamma_5  e_j \cdot  \psi \otimes [ \nabla_{e_i}^{\mathcal{H}_f} , \Phi] \, [ \nabla_{e_j}^{\mathcal{H}_f} , \Phi] 
\chi
+ \psi \otimes (\Phi^4) \, \chi - 2 \, E \, \psi \otimes (\Phi^2) \, \chi
\nonumber \\
&&+ \sum_ {i\ne j} \left( E \,e_i\cdot e_j\cdot\psi\right)
\otimes\left(\Omega^{\mathcal{H}}_{ij}\;\chi \right)
+ \sum_{i=1}^n \left( E \,  \gamma_5  e_i  \cdot   \psi 
+  \gamma_5  e_i \cdot  E \, \psi \right) 
\otimes [ \nabla_{e_i}^{\mathcal{H}_f} , \Phi] \,  \chi
\nonumber \\
&&+ \frac{1}{2}  \cdot\sum_ {{i \ne j} \atop{k}}
\left( e_i\cdot e_j\cdot \gamma_5 e_k \cdot  \psi \right)\otimes
\left(\Omega^{\mathcal{H}_f}_{ij} 
\,  [ \nabla_{e_k}^{\mathcal{H}_f} , \Phi]   \;\chi \right) 
\nonumber \\
&&+ \frac{1}{2}  \cdot\sum_ {{i \ne j} \atop{k}}
\left(  \gamma_5 e_k \cdot e_i\cdot e_j\cdot \psi \right)\otimes
\left(  [ \nabla_{e_k}^{\mathcal{H}_f} , \Phi]  \, \Omega^{\mathcal{H}_f}_{ij} \;\chi \right) 
\nonumber \\
&&- \frac{1}{2}  \cdot\sum_ {i \ne j}
\left( e_i\cdot e_j\cdot   \psi \right)\otimes
\left((\Omega^{\mathcal{H}_f}_{ij} 
\,  (\Phi^2) +  (\Phi^2) \,  \Omega^{\mathcal{H}_f}_{ij} )  \;\chi \right) 
\nonumber \\
&&-  \sum_{i=1}^n  \gamma_5  e_i \cdot   \psi \otimes 
\left( (\Phi^2) \, [ \nabla_{e_i}^{\mathcal{H}_f} , \Phi] 
+ [ \nabla_{e_i}^{\mathcal{H}_f} , \Phi]  \, (\Phi^2) \right) \, \chi.
\label{EPhi2}
\ee
Only the first five summands on the right-hand side contribute to the
trace of $(E_\Phi)^2$.
In four dimensions $E$ consists of two summands proportional  to
the identity and one summand proportional to  $\gamma_5$.
Therefore the endomorphism defined by the sixth and seventh summand 
of (\ref{EPhi2})
\bb
\psi \otimes \chi \mapsto 
 \sum_ {i\ne j} \left( E \,e_i\cdot e_j\cdot\psi\right)
\otimes\left(\Omega^{\mathcal{H}}_{ij}\;\chi \right)
+ \sum_{i=1}^n \left( E \,  \gamma_5  e_i  \cdot   \psi 
+  \gamma_5  e_i \cdot  E \, \psi \right) 
\otimes [ \nabla_{e_i}^{\mathcal{H}_f} , \Phi] \,  \chi
\nonumber 
\ee
is traceless due to Clifford relations and cyclicity of
the trace. The trace of the endomorphism given by the remaining summands
of (\ref{EPhi2}) vanishes due to Clifford relations without
employing that the dimension of the manifold is four (in other
dimensions  $\gamma_5$ is then the volume element).
Thus we find for the trace
\bb
\ttt (E_\Phi^2) = \rank \mathcal{H}_f \cdot
\tr (E^2) + \tr_f  (\Omega^{\mathcal{H}_f}_{ij} \Omega^{\mathcal{H}_f}_{ij})
+ 4 \cdot \tr_f ([\nabla^{\mathcal{H}_f}, \Phi]^2) + 4 \cdot
\tr_f (\Phi^4) - 2 \cdot  \tr (E) \cdot \tr_f (\Phi^2).
\label{EPhi2tr}
\ee
The last ingredient we need to calculate $a_4(D_\Phi^2)$ is the trace the squared
curvature tensor $ \Omega^{\overline{\nabla}}_{ij}$, see (\ref{tensorcurvature}), 
associated to the connection
in the Bochner formula (\ref{bochnertwist}). We note that  
$\ttt ( \Omega_{ij} \otimes  \Omega^{\mathcal{H}_f}_{ij}) = 0$ and therefore we find
\bb
\sum_{i,  j} \ttt \,\left(  \Omega^{\overline{\nabla}}_{ij} \, \Omega^{\overline{\nabla}}_{ij} 
\right) &=& \sum_{i,  j} \ttt \, \left( (\Omega_{ij} \Omega_{ij}) \otimes 1_{\mathcal{H}_f} 
+1_4 \otimes (\Omega^{\mathcal{H}_f}_{ij} \Omega^{\mathcal{H}_f}_{ij})
+ 2  \,  \Omega_{ij} \otimes  \Omega^{\mathcal{H}_f}_{ij} \right)
\nonumber \\
&=& \rank \, \mathcal{H}_f \cdot  \sum_{i,  j} \tr \, \left(\Omega_{ij} \Omega_{ij} \right)
+ 4 \cdot \sum_{i,  j} \tr_f  \, \left( \Omega^{\mathcal{H}_f}_{ij} 
\Omega^{\mathcal{H}_f}_{ij} \right).
\nonumber
\ee 
We choose the finite space $\mathcal{H}_f$
according to the construction of the noncommutative Standard Model 
\cite{cc,con,mcc}, i.e. $\rank \mathcal{H}_f = 96$ and $\nabla^{\mathcal{H}_f}$
is the appropriate  covariant derivative associated to the Standard Model gauge group
$U(1)_Y \times SU(2)_w \times SU(3)_c$.

Inserting  the above results into the spectral action (\ref{specact}) we obtain
for the bosonic Lagrangian of the Standard Model coupled to gravity and torsion:
\bb
I_{bos.} = \ttt \, F \left( \frac{D_\Phi^2}{\Lambda^2} \right) &=& 
\Lambda^4 \, F_4 \, a_0 (D_\Phi^2) + \Lambda^2 \, F_2 \, a_2(D_\Phi^2)  
+ \Lambda^0 \, F_0 \, a_4(D_\Phi^2)  +  \mathrm{O} (\Lambda^{-2})
\nonumber \\
&=& \frac{\Lambda^4 \, F_4}{16 \pi^2} \int_M \ttt ( 1_{\rank \mathcal{H}}) \, dvol
\nonumber \\
&&+ \frac{\Lambda^2 \, F_2}{96 \pi^2} \int_M  \ttt \left( 6 E_\Phi + R \right) \, dvol
\nonumber \\
&& + \frac{ \Lambda^0 \, F_0}{5760 \pi^2} \int_M 
\Big\{  \ttt \left( 60 R E_\Phi + 180 E_\Phi^2 +
30 \, \Omega^{\overline{\nabla}}_{ij} \Omega^{\overline{\nabla}}_{ij}\right) 
 \nonumber \\
&& \qquad \qquad \quad + 20 R^2 - 8 \|Ric\|^2 + 8 \|Riem\|^2 \Big\} \, dvol 
+  \mathrm{O} (\Lambda^{-2})
\nonumber \\
\nonumber \\
&=& \frac{24 \Lambda^4 \, F_4}{ \pi^2} \int_M  \, dvol
\nonumber \\
&&+ \frac{\Lambda^2 \, F_2}{96 \pi^2} \int_M  \left\{ 
96  \big( 6 \tr(E)  +  4 R  \big)  - 4 \tr_f (\Phi^2)
\right\} \, dvol 
\nonumber \\
&& + \frac{  F_0}{5760 \pi^2} \int_M 
\Big\{  96 \, \big( 60 R \tr(E) + 180 \tr (E^2) +
30 \tr( \Omega_{ij} \Omega_{ij})
 \nonumber \\
&& \qquad \qquad \qquad + 80 R^2 - 32 \|Ric\|^2 + 32 \|Riem\|^2 \big) 
\nonumber \\
&& \qquad \qquad \qquad + 300 \, \tr_f ( \Omega^{\mathcal{H}_f}_{ij}  \Omega^{\mathcal{H}_f}_{ij})  
+ 720\, \tr_f ([\nabla^{\mathcal{H}_f}, \Phi]^2) 
\nonumber \\
&& \qquad \qquad \qquad  + 720 \,
\tr_f (\Phi^4) - 360 \,  \tr (E) \cdot \tr_f (\Phi^2)
\Big\} \, dvol +  \mathrm{O} (\Lambda^{-2})
\nonumber
\ee
To cast this action into a more familiar form we use the standard formulas to 
express $ \tr_f ( \Omega^{\mathcal{H}_f}_{ij}  \Omega^{\mathcal{H}_f}_{ij}) $
in terms of the norms of the gauge field strengths $ \| G \|^2 = \sum_{\mu,\nu,i} G^i_{\mu \nu} G^{\mu \nu i}$, $ \| F \|^2 = \sum_{\mu, \nu, \alpha} F^\alpha_{\mu \nu}
F^{\mu \nu \alpha} $, $ \| B \|^2 = \sum_{\mu \nu}  B_{\mu \nu} B^{\mu \nu}$.
Here $G^i_{\mu \nu} $ is the curvature of the $SU(3)_c$-connection 
with coupling $g_3$, $ F^\alpha_{\mu \nu} $ is the curvature of the $SU(2)_w$-
connection  with coupling $g_2$ and $ B_{\mu \nu} $ is the curvature of the 
$U(1)_Y$-connection  with coupling $g_1$.

We also calculate the traces of of the Higgs endomorphisms $\Phi^2$ and
$\Phi^4$ explicitly in terms of the Higgs doublet $\varphi$ and obtain
$\tr_f (\Phi^2) =  4 a \, | \varphi |^2 + 2 c$ and
$\tr_f (\Phi^4) = 4 b \, | \varphi |^4 + 8 e  | \varphi |^2 + 2 d$. The coefficients
$a$, $b$, $c$, $d$ and $e$ are traces of the $3\times3$ Yukawa matrices for
the quarks ($k_u$ and $k_d$), the leptons ($k_e$ and $k_\nu$) and 
the Majorana mass matrix  for the right-handed  neutrinos ($k_{\nu_R}$)
given by
\bb
a&=& \tr_3 ( 3 |k_u|^2 + 3 |k_d|^2 + |k_e|^2 + |k_\nu|^2 ),
\nonumber \\
b&=& \tr_3 ( 3 |k_u|^4 + 3 |k_d|^4 + |k_e|^4 + |k_\nu|^4 ),
\nonumber \\
c& =& \tr_3 ( |k_{\nu_R}|^2 ),
\nonumber \\
d &= &\tr_3 ( |k_{\nu_R}|^4 ),
\nonumber \\
e& =& \tr_3 ( |k_\nu|^2 |k_{\nu_R}|^2 ).
\nonumber 
\ee
We conclude that the spectral action principle predicts the following
form of the bosonic Lagrangian for the Standard model in the presence
of skew-symmetric torsion:
\bb
I_{bos.} 
&=& \frac{ 24 \Lambda^4 F_4 }{\pi^2}  \int_M   dvol
\nonumber \\
&& 
+ \frac{\Lambda^2 F_2}{ \pi^2}  \int_M     \left\{ 27 \, T_0^2 - 2 R 
- a |\varphi|^2 - \frac{1}{2} c \right\} dvol 
\nonumber \\
&& + \frac{F_0}{2 \pi^2} \int_M \Big\{  
 \frac{1}{6} R^2 - \frac{4}{15} \|Ric\|^2
-  \frac{7}{30} \|Riem\|^2  + \frac{243}{8}
(T_0^2)^2  
 - \frac{27}{4} \| c(T) \|^2
 \nonumber \\
&&  \qquad \qquad \quad   
+ \frac{27}{4} \|dT\|^2 - \frac{9}{2} \| d^*T\|^2
- \frac{9}{2} \|sym_0^2 (\nabla T)\|^2   - \frac{9}{2} R(T)
\nonumber \\
&&  \qquad \qquad \quad   
+ g_3^2  \| G \|^2  + g_2^2  \| F \|^2  + \frac{5}{3} g_1^2 \| B \|^2
 \nonumber \\
&&  \qquad \qquad \quad 
+ a |D_\nu \varphi|^2 + b |\varphi|^4 + 2 e |\varphi|^2 + \frac{1}{2} d
+ \frac{1}{6} R \, \left( a |\varphi|^2 +  \frac{1}{2} c \right)
\nonumber \\
&&  \qquad \qquad \quad 
- \frac{9}{4} T_0^2 \, \left( a |\varphi|^2 +  \frac{1}{2} c \right) 
\Big\} dvol +  \mathrm{O} (\Lambda^{-2}) .
\label{final}
\ee
Here  we were able to use the standard results from the torsion free
case, see \cite[p.22]{cc2} or \cite{mcc,iks}.
As in the pure gravity+torsion case the torsion becomes dynamical and 
couples only with the trace free part of the Riemann curvature tensor.

In presence of the Standard Model fields we obtained essentially one new
term (apart from the usual suspects) coupling the torsion to the 
Higgs field
\bb
I_{new} =  - \frac{ 9 a F_0}{8 \pi^2} \int_M   T_0^2   |\varphi|^2  \, dvol.
\ee
This is another amazing feature of the spectral action principle: it 
supports the interpretation of the Higgs field as the gravitational field of the internal space in the noncommutative product geometry \cite{con}.
The full Standard Model action is given by
\bb
I_{SM} = \ttt F \left( \frac{D_\Phi^2}{\Lambda^2} \right) + \frac{1}{2} \langle J \Psi , D_\Phi \Psi \rangle
\qquad {\rm with} \quad \Psi \in \mathcal{H}
\ee
where the fermionic action $\frac{1}{2} \langle J \Psi , D_\Phi \Psi \rangle$ contains a 
coupling between torsion and the fermions, and $J$ is the real structure of
the spectral triple. This action takes care of the fermion doubling problem, compare
\cite[(5.9)]{cc2}.

\subsection*{Conclusions}

We have calculated the spectral action for Dirac operators arising from
geometries with skew-symmetric torsion and their twisted version originating
in the noncommutative approach to the standard model. 
In both cases we find that torsion couples to the trace free part of the 
Riemann curvature tensor 
and in the latter case to the Higgs boson of the standard model.
Furthermore the torsion becomes dynamical due to derivative terms in 
the action.

Now one certainly has to wonder about possible experimental signatures
of these new phenomena, both on local scales (Earth and the solar system) and
cosmological scales. 

We may assume for the moment that the Schwarzschild metric is a good approximation to
the gravitational field of the Earth (even for the ``new'' spectral action with
torsion). The Weyl curvature  of the Schwarzschild metric is non-zero 
and hence torsion for the corresponding critical point of the action seems
probable.  
Then we would  expect  effects on freely falling particles or atoms with 
different spins.
This might lead to measurable effects in atom interferometry experiments.

The cosmological consequences are much more speculative.
It has been noted  \cite{rovelli,mercuri} that torsion induces 
four-fermion interactions which in turn may provide a possible solution
to the problem of the enormously
large cosmological constant $\Lambda_c \sim \Lambda \sim 10^{17}$GeV predicted by the  spectral action (without torsion) \cite{condensate}. In this 
framework also a natural mechanism for inflation appears naturally.

To obtain more rigourous results it will be necessary to investigate the
Euler-Lagrange equations of the spectral action with torsion. It would be
interesting to find exact solutions with non-vanishing torsion and compare
them with the  known solutions of Einstein's equations.

\subsection*{Appendix: Bochner formula for twisted Dirac operators in presence of torsion}

We consider a closed Riemannian spin manifold $M$ of dimension $n$ and a 
connection $\nabla$ on the tangent bundle which is compatible with the metric 
and has totally anti-symmetric torsion $T$.
By $\nabla$ we also denote the induced connection on the spinor bundle $S$ 
of $M$ which can be expressed as in (\ref{tconnection}).
Given a vector bundle $\mathcal{H}$ over $M$ with connection $\nabla^{\mathcal{H}}$ 
we consider the tensor connection $\widehat{\nabla}=\nabla \otimes \id_{\mathcal{H}}
+ \id_{S} \otimes \nabla^{\mathcal{H}}$ and the associated twisted Dirac operator 
$D^{\widehat{\nabla}}$ acting on sections of $S\otimes\mathcal{H}$.
We define another tensor connection
\bb
\overline{\nabla}=\widetilde{\nabla} \otimes \id_{\mathcal{H}}
+ \id_{S} \otimes \nabla^{\mathcal{H}}, 
\label{overline}
\ee
where $\widetilde{\nabla}$ is the spin connection
from (\ref{drittelconn}), and we claim the Bochner formula
\bb
\left(D^{\widehat{\nabla}}\right)^2=\Delta^{\overline{\nabla}}- \widetilde E,
\label{bochnertwist}
\ee
where $\Delta^{\overline{\nabla}}$ denotes the horizontal Laplacian associated to 
$\overline{\nabla}$ and $\widetilde  E$ denotes an endomorphism field of $S\otimes\mathcal{H}$
which still has to be determined.

To that end we fix an arbitrary point $p\in M$, and we choose a local orthonormal basis of
vector fields $e_1,\ldots,e_n$ with $\nabla e_i=0$ in $p$ (for any $i$).
From (\ref{zero}) we get that $\nabla^{LC}_{e_i}e_i=\nabla_{e_i}e_i=0$ in $p$ (for any $i$)
and the Lie bracket $[e_i,e_j] = \nabla^{LC}_{e_i}e_j-\nabla^{LC}_{e_j}e_i=-T(e_i,e_j,\cdot)^\#$  
in $p$ (for any $i,j$).
For any section $\psi$ of $S$ and any section $\chi$ of $\mathcal{H}$ we get in $p$:
\bb
D^{\widehat{\nabla}}\left(\psi\otimes\chi\right)&=&
\sum_{i=1}^n\left\{\left(e_i\cdot \nabla_{e_i}\psi \right)\otimes\chi 
+ \left(e_i\cdot\psi \right)\otimes \nabla^{\mathcal{H}}_{e_i}\chi\right\},
\label{DPhi} \\
\left(D^{\widehat{\nabla}}\right)^2\left(\psi\otimes\chi\right)&=&
\left( D^2\psi\right)\otimes\chi
+\sum_{i,j=1}^n\left( e_j\cdot e_i\cdot\psi\right)
\otimes \nabla^{\mathcal{H}}_{e_j}\nabla^{\mathcal{H}}_{e_i}\chi
\nonumber \\
&&\qquad
+\sum_{i,j=1}^n\left\{ \left( e_j\cdot e_i\cdot\nabla_{e_i}\psi\right)\otimes \nabla^{\mathcal{H}}_{e_j}\chi
+\left( e_j\cdot e_i\cdot\nabla_{e_j}\psi\right)\otimes \nabla^{\mathcal{H}}_{e_i}\chi
\right\}
\nonumber \\
&=&\left( D^2\psi\right)\otimes\chi
+\sum_{i,j=1}^n\left( e_j\cdot e_i\cdot\psi\right)
\otimes \nabla^{\mathcal{H}}_{e_j}\nabla^{\mathcal{H}}_{e_i}\chi
-2\cdot\sum_{i=1}^n \nabla_{e_i}\psi\otimes \nabla^{\mathcal{H}}_{e_i}\chi,
\nonumber
\ee
where the last equation holds due to Clifford relations.

As before,  let $\Delta$ denote the Laplacian associated to the spin connection $\widetilde{\nabla}$.
In $p$ we obtain for the Laplacian associated to $\overline{\nabla}$:
\bb
\Delta^{\overline{\nabla}}\left(\psi\otimes\chi\right)&=& \left(\Delta\psi\right)\otimes\chi
-\sum_{i=1}^n \psi\otimes\nabla^{\mathcal{H}}_{e_i}\nabla^{\mathcal{H}}_{e_i}\chi
-2\cdot\sum_{i=1}^n \widetilde{\nabla}_{e_i}\psi\otimes \nabla^{\mathcal{H}}_{e_i}\chi
\nonumber \\
&=& \left(\Delta\psi\right)\otimes\chi
+\sum_{i=1}^n e_i\cdot e_i\cdot \psi\otimes\nabla^{\mathcal{H}}_{e_i}\nabla^{\mathcal{H}}_{e_i}\chi
-2\cdot\sum_{i=1}^n\nabla_{e_i}\psi\otimes \nabla^{\mathcal{H}}_{e_i}\chi
-\sum_{i=1}^n \left((e_i \lrcorner T)\cdot\psi\right)\otimes  \nabla^{\mathcal{H}}_{e_i}\chi,
\nonumber
\ee
where we have used (\ref{tconnection}) and (\ref{drittelconn}).
In $p$ we observe
\bb
\sum_{i,j=1}^n \left(e_i\cdot e_j\cdot\psi\right)\otimes\nabla^{\mathcal{H}}_{[e_i,e_j]}\chi
&=&-\sum_{i,j=1}^n\left(e_i\cdot e_j\cdot\psi\right)\otimes\nabla^{\mathcal{H}}_{T(e_i,e_j,\cdot)^\#}\chi
=-\sum_{i,j,k=1}^n \left(e_i\cdot e_j\cdot\psi\right)\otimes\nabla^{\mathcal{H}}_{T(e_i,e_j,e_k)e_k}\chi
\nonumber \\
&=&-\sum_{i,j,k=1}^n \left(T(e_i,e_j,e_k)\,e_i\cdot e_j\cdot\psi\right)\otimes\nabla^{\mathcal{H}}_{e_k}\chi
=-2\cdot \sum_{k=1}^n \left((e_k \lrcorner T)\cdot\psi\right) \otimes  \nabla^{\mathcal{H}}_{e_k}\chi.
\nonumber
\ee
Putting all this together we get in $p$ that
\bb
\left(D^{\widehat{\nabla}}\right)^2\left(\psi\otimes\chi\right)
-\Delta^{\overline{\nabla}}\left(\psi\otimes\chi\right)&=&
\left( D^2\psi-\Delta\psi \right)\otimes\chi\nonumber \\
&& \quad
+ \sum_{i\ne j}\left( e_i\cdot e_j\cdot\psi\right)
\otimes \nabla^{\mathcal{H}}_{e_i}\nabla^{\mathcal{H}}_{e_j}\chi
- \frac{1}{2}\cdot\sum_ {i\ne j}\left(e_i\cdot e_j\cdot\psi\right)\otimes\nabla^{\mathcal{H}}_{[e_i,e_j]}\chi\nonumber \\
&=& \left( D^2\psi-\Delta\psi \right)\otimes\chi 
- \frac{1}{2}\cdot\sum_ {i\ne j}\left(e_i\cdot e_j\cdot\psi\right)\otimes\left({\Omega^{\mathcal{H}}}_{ij}\;\chi \right),
\nonumber
\ee
where $\Omega^{\mathcal{H}}_{ij}=\nabla^{\mathcal{H}}_{e_i}\nabla^{\mathcal{H}}_{e_j}-
\nabla^{\mathcal{H}}_{e_j}\nabla^{\mathcal{H}}_{e_i}-\nabla^{\mathcal{H}}_{[e_i,e_j]}$ is the curvature 
endomorphism of the twist bundle $\mathcal{H}$.
Taking (\ref{DD}) into account we can identify the endomorphism field $\widetilde  E$ in the following Bochner formula 
(\ref{bochnertwist}) as
\bb
\label{twistpotential}
\widetilde  E\left(\psi\otimes\chi\right) &=&
\left((-\frac{3}{4} dT - \frac{1}{4} R + \frac{9}{8} T_0^2)\,\psi \right)\otimes\chi
+ \frac{1}{2} \cdot\sum_ {i\ne j} \left(e_i\cdot e_j\cdot\psi\right)\otimes\left(\Omega^{\mathcal{H}}_{ij}\;\chi \right)
\nonumber \\
&=&
\left(E \,\psi \right)\otimes\chi
+ \frac{1}{2} \cdot\sum_ {i\ne j} \left(e_i\cdot e_j\cdot\psi\right)\otimes\left(\Omega^{\mathcal{H}}_{ij}\;\chi \right).
\label{Endo}
\ee
Finally we remark that the curvature of the connection $\overline{\nabla}=\widetilde{\nabla} \otimes \id_{\mathcal{H}}
+ \id_{S} \otimes \nabla^{\mathcal{H}}$ is given by
\begin{equation}
\label{tensorcurvature}
{\Omega^{\overline{\nabla}}}_{ij}\left(\psi\otimes\chi\right)
=\left(\Omega_{ij}\psi\right)\otimes\chi
+\psi\otimes\left(\Omega^{\mathcal{H}}_{ij}\;\chi\right),
\end{equation}
where $\Omega_{ij}$ is the curvature of the spin connection $\widetilde{\nabla}$.

\subsection*{Acknowledgements}

We gratefully acknowledge funding of this work
by the Deutsche Forschungsgemeinschaft in
particular the SFB 647 ``Raum-Zeit-Materie''.

\vfil\eject
\enlargethispage{1cm}
\thispagestyle{empty}


\begin{thebibliography}{50}

\bibitem{con}
 A. Connes, {\it Noncommutative Geometry}, Academic Press (1994)\\
 A. Connes, {\it Noncommutative geometry and
reality},  J. Math. Phys. 36 (1995) 6194\\
A. Connes, {\it Gravity coupled with matter and the
foundation of noncommutative geometry}, hep-th/9603053, Comm.
Math. Phys. 155 (1996) 109\\
A. Connes \& M. Marcolli {\it Noncommutative Geometry, Quantum Fields and Motives }
(2007) http://www.alainconnes.org/ 

\bibitem{cc}
 A. Chamseddine \& A. Connes, {\it The spectral action principle},
hep-th/9606001, Comm. Math. Phys. 182 (1996) 155


\bibitem{mcc}
A. Chamseddine, A. Connes \& M. Marcolli,
{\it Gravity and the standard model with neutrino mixing},
hep-th/0610241, Adv.Theor.Math.Phys.11:991 (2007), \\
A. Chamseddine \& A. Connes,
{\it Why the Standard Model},
arXiv:0706.3688 [hep-th], J.Geom.Phys.58:38 (2008), \\
A. Chamseddine \& A. Connes,
{\it Conceptual Explanation for the Algebra in the Noncommutative Approach to the Standard Model},
arXiv:0706.3690 [hep-th], Phys.Rev.Lett.99:191601 (2007)


\bibitem{ccdil}
A. Chamseddine \& A. Connes,
{\it  Scale invariance in the spectral action},
hep-th/0512169, J.Math.Phys.47:063504, (2006)

\bibitem{ccbound}
A. Chamseddine \& A. Connes,
{\it  Quantum Gravity Boundary Terms from Spectral Action},
arXiv:0705.1786 [hep-th],  Phys.Rev.Lett.99:071302 (2007) 

\bibitem{af}
I. Agricola \& T. Friedrich, {\it On the holonomy of connections with
skew-symmetric torsion}, Math. Ann. 328 (2004) 711

\bibitem{cc2}
A. Chamseddine \& A. Connes {\it Noncommutative Geometry as a Framework for Unification of all Fundamental Interactions including Gravity. Part I},
arXiv:1004.0464 [hep-th]

\bibitem{gilk}
P.B. Gilkey, {\it Invariance Theory, the Heat Equation, and the Atiyah-Singer Index
Theorem (second edition)}, CRC Press, Boca Raton (1995)  

\bibitem{nvw} 
R. Nest, E. Vogt \& W. Werner, {\it Spectral Action and the Connes-Chamseddine
Model}, in F. Scheck, H. Upmeier \& W. Werner (eds.), Lecture Notes in Physics
596 (2002) 109

\bibitem{OBU}
W.H. Goldthorpe,
{\it Spectral Geometry and SO(4) Gravity in a Riemann-Cartan Spacetime},
Nucl.Phys. B170:307 (1980)
\\
Yu.N. Obukhov,
{\it Spectral Geometry Of The Riemann-Cartan Space-Time},
Nucl.Phys.B212:237 (1983) 
\\
I.L. Buchbinder, S.D. Odintsov \& I.L. Shapiro,
{\it Nonsingular cosmological model with torsion induced by vacuum
quantum effects},
Phys. Lett.B162:92 (1985)
\\
G. Grensing,
{\it  Induced Gravity For Nonzero Torsion},
Phys.Lett.B169:333 (1986)
\\
G. Cognola \& S. Zerbini,
{\it Seeley-deWitt Coefficents in a Riemann-Cartan Manifold},
Phys.Lett.B214:70 (1988)

\bibitem{SAL}
S. Salamon,
{\it Riemannian geometry and holonomy groups},
Pitman Research Notes in Mathematics Series, 201;  
Wiley \& Sons, Inc., New York, 1989

\bibitem{KW}
W. Kalau \& M. Walze,
{\it Gravity, noncommutative geometry and the Wodzicki residue},
gr-qc/9312031, J.Geom.Phys.16:327 (1995)

\bibitem{AT}
T. Ackermann \& J. Tolksdorf,
{\it  The Generalized Lichnerowicz formula and analysis of Dirac operators},
preprint: CPT-95-P-3166, MANNHEIMER-186, Jan 1995, hep-th/9503153

\bibitem{iks}
B. Iochum, D. Kastler \& T. Sch\"ucker,
{\it On the universal Chamseddine-Connes action. I. Details of the action
computation},
J.Math.Phys. 38 (10): 4929 (1997)

\bibitem{rovelli}
A. Perez \& C. Rovelli,
{\it Physical effects of the Immirzi parameter},
gr-qc/0505081, Phys.Rev.D73:044013 (2006)

\bibitem{mercuri}
S. Mercuri, {\it Fermions in the Ashtekar-Barbero connection formalism
for arbitrary values of the Immirzi parameter}, 
Phys.Rev.D73:084016 (2006)

\bibitem{condensate}
S. Alexander \& D. Vaid, 
{\it Gravity induced chiral condensate formation and the cosmological constant},
e-Print: hep-th/0609066 (2006)
\\
S. Alexander \& D. Vaid, 
{\it A Fine tuning free resolution of the cosmological constant problem}
e-Print: hep-th/0702064 (2007)
\\
S. Alexander \& T. Biswas,  
{\it  The Cosmological BCS mechanism and the Big Bang Singularity},
arXiv:0807.4468 [hep-th]  Phys.Rev.D80:023501 (2009)

\end{thebibliography}
\end{document}